\documentclass[a4paper,article,10pt]{memoir}
\usepackage[english]{babel}
\usepackage{CCLAuthor}
\usepackage[reqno]{amsmath}
\usepackage{amssymb}
\usepackage{amsthm}  
\theoremstyle{plain}

\theoremstyle{definition}

\numberwithin{theorem}{chapter}
\usepackage[round,authoryear]{natbib} 
\usepackage{graphicx}                 
\graphicspath{{figs/}}
\usepackage{bm}
\usepackage[mathcal]{euscript}
\usepackage{psfrag}
\usepackage{subfigure}
\newcommand{\ie}{\textit{i.e.}}
\newcommand{\etal}{et~al.} 
\newcommand{\mathnotation}[2]{\newcommand{#1}{\ensuremath{#2}}}
\DeclareMathOperator{\trace}{tr}		

\renewcommand{\l}{\left}			
\renewcommand{\r}{\right}			
\mathnotation{\ldef}{\mathrel{\raisebox{.069ex}{:}\!\!=}}
\mathnotation{\rdef}{\mathrel{=\!\!\raisebox{.069ex}{:}}}
\mathnotation{\nn}{N}				
\mathnotation{\htop}{h_{\mathrm{top}}}		

\renewcommand{\time}{t}
\mathnotation{\xc}{x}
\mathnotation{\yc}{y}
\mathnotation{\xv}{\bm{x}}
\mathnotation{\xci}{\xc_\iter}
\mathnotation{\yci}{\yc_\iter}
\mathnotation{\xcii}{\xc_{\iter+1}}
\mathnotation{\ycii}{\yc_{\iter+1}}
\mathnotation{\Xc}{X}
\mathnotation{\Yc}{Y}
\mathnotation{\Xv}{\bm{X}}
\mathnotation{\tXc}{\widetilde{X}}
\mathnotation{\tYc}{\widetilde{Y}}
\mathnotation{\tXv}{\widetilde{\bm{X}}}
\mathnotation{\velc}{u}
\mathnotation{\velv}{\bm{\velc}}
\mathnotation{\iter}{n}
\mathnotation{\qfpar}{\alpha}
\mathnotation{\Jac}{\mathcal{J}}
\mathnotation{\Yci}{\Yc_{\mathrm{inv}}}

\begin{document}
%
%
%
\title{The Size of Ghost Rods}
%
%
\author{%
    Jean-Luc Thiffeault, 
    Emmanuelle Gouillart,
    and 
    Matthew D. Finn 
    \\ \smallskip\small
    Department of Mathematics, Imperial College
    London, SW7 2AZ, United Kingdom
    }
    \maketitle
%
%
%
    \vskip-2\baselineskip\noindent \begin{abstract}
      `Ghost Rods' are periodic structures in a two-dimensional flow that have
      an effect on material lines that is similar to real stirring rods.  An
      example is a periodic island: material lines exterior to it must wrap
      around such an island, because determinism forbids them from crossing
      through it.  Hence, islands act as topological obstacles to material
      lines, just like physical rods, and lower bounds on the rate of
      stretching of material lines can be deduced from the motion of islands
      and rods.  Here, we show that unstable periodic orbits can also act as
      ghost rods, as long as material lines can `fold' around the orbit, which
      requires the orbit to be parabolic.  We investigate the factors that
      determine the effective size of ghost rods, that is, the magnitude of
      their impact on material lines.
    \end{abstract}

\bigskip

\noindent
\textsl{To be published in `Proceedings of the Workshop on Analysis and
Control of Mixing with Applications to Micro and Macro Flow Processes,' CISM,
Udine, Italy, July 2005. (Springer-Verlag, 2006)}

\CCLsection{Introduction}
\label{sec:intro}

Topological kinematics is the application of topology to chaotic advection in
fluids.  In two dimensions, braids are the natural mathematical construct to
use for a topological analysis.  Boyland~\etal\ used braids very effectively
to analyse the motion of stirring rods in viscous flow \citep{Boyland2000} and
point vortices in ideal flow \citep{Boyland2003}.  A braid is associated with
the motion of the rods or vortices by plotting their trajectory in a
space-time diagram: the resulting ``spaghetti plot'' is obviously a braid.
Here, we shall not be too concerned with the precise mathematical properties
of braids---the intuitive, capillary notion of what a braid resembles will
suffice.

Rods and points vortices share the common feature that they are topological
obstacles to material lines in two dimensions.  Of course, any fluid particle
is such an obstacle, and recently one of us analysed braids formed by particle
trajectories \citep{Thiffeault2005}.  The fact that particle orbits are
topological obstacles puts a lower bound on the \emph{topological
entropy}---the growth rate of material lines~\citep{Boyland2000,Newhouse1993}.
Imagine a material line that is initially linked around the topological
obstacles under consideration (rods or fluid particles).  Then as the position
of these obstacles evolves the material line is dragged along, and as the
particles braid around each other the material line must grow by at least a
certain amount.  The properties of the braid thus imply a lower bound on the
growth rate of material lines in the fluid.

\begin{figure}[htbp]
\centering
\includegraphics[width=.5\textwidth]{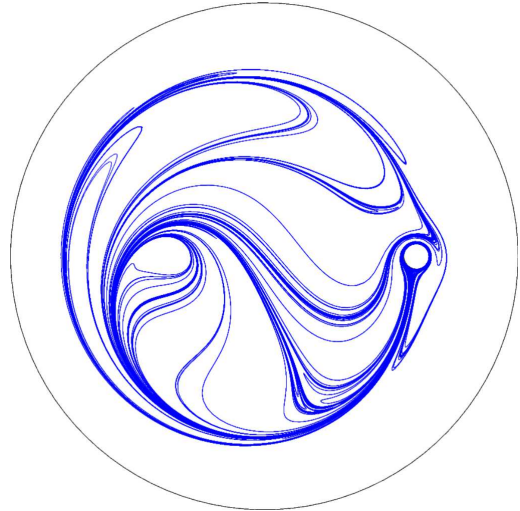}
\caption{A material line being stirred by a moving rod in a viscous fluid.
  The rod is the circle visible in the centre-right portion of the fluid, but
  observe that there is a rod-like structure in the centre-left portion.  This
  is a periodic island that acts like a rod---a ghost rod.}
\label{fig:ghostrod}
\end{figure}

For time-periodic flows, the natural topological obstacles to study are fluid
parcels associated with particular periodic orbits.  Recently, we introduced
to fluid mechanics the concept of ``ghost rods''
\citep{Gouillart2005preprint}.  We analysed the motion of material rods,
stable periodic orbits (islands), and unstable periodic orbits from a
topological perspective.  We showed that periodic orbits associated with
islands behave very similarly to material rods: they are large topological
obstacles ploughing through the fluid, and material lines must get out of
their way or else wrap around them.  Periodic islands have the advantage of
being easily identifiable visually, and can clearly be regarded as ``rods.''
Figure~\ref{fig:ghostrod} illustrates this: there is only one physical rod
stirring the flow, but a ghostly second rod is clearly visible in the
centre-left portion of the plot, around which material lines are wrapped.
Indeed, there is a regular island in that region
\citep{Gouillart2005preprint}.

The foregoing description is topological in nature.  The size of the rods is
immaterial to the topological entropy \citep{Boyland2000,MattFinn2003}.  Of
course, in practice their size matters a lot: if the rods are made smaller, so
does the region of the fluid for which the topological entropy lower bound
applies.  In the limit of infinitely small rods, one might expect this region
to shrink to zero.  This is certainly true of physical rods: if they were made
infinitely small, the fluid would not even notice their presence and nothing
would happen, except in a vanishingly small region.  There is currently no
theory that gives the size of the affected region given the size of the rods
and their path, but in practice it is observed (in viscous flows) that it is of
the order of the size of the rods and the size of the region swept by their
motion.

So much for physical rods.  But what about ghost rods?  As their name
indicates, they have no material existence.  However, since they behave much
like physical rods, we may ask what is their effective size.  That is,
topologically a ghost rod is supposed to mimic a real physical rod, but how
much of an impact does it effectively have on the surrounding fluid?  For
periodic islands, the answer is clear: the effective size of the ghost rod is
the size of the island.  Figure~\ref{fig:ghostrod} convincingly illustrates
that, as far as material lines are concerned, there is a stirring rod of the
size of the periodic island in the centre-left portion of the flow.

For unstable periodic orbits, the answer is much less clear, since in
principle ghost rods of this type have zero size.  In this paper, we shall
investigate the effective size of ghost rods associated with unstable periodic
orbits.  In fact, as we shall see, not all unstable periodic orbits can even
be said to be ghost rods.  Rather, only unstable periodic orbits of parabolic
(as opposed to hyperbolic) type can hope to qualify as ghost rods.  The local
linear structure near an hyperbolic orbit prevents material lines from
``wrapping'' around the periodic point, so that it does not appear as a rod at
all.  For parabolic orbits of a certain type, the unstable manifold terminates
at the periodic orbit, allowing material lines to wrap around the point
without encountering the invariant manifold.  Thus, the periodic orbit appears
\emph{visually} as a tiny rod, which is our criterion for considering periodic
orbits to be ghost rods.

\CCLsection{Unstable Periodic Orbits}

\begin{figure}[htbp]
\psfrag{unstable}{unstable}
\psfrag{stable}{stable}
\centering
\includegraphics[width=.3\textwidth]{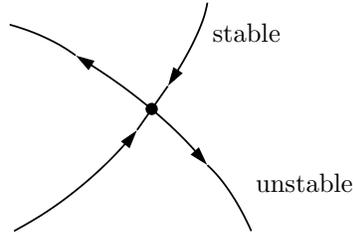}
\caption{The stable and unstable manifolds of a hyperbolic point.}
\label{fig:hyper}
\end{figure}

In an incompressible flow, the linearised flow around an unstable periodic
orbit can be one of two types.  Figure~\ref{fig:hyper} depicts the most
common, called a hyperbolic orbit, or hyperbolic point if one is speaking of
the Poincar\'e section (stroboscopic map) of the time-periodic flow.
There are two distinguished directions along which points respectively get
away from or converge to the periodic orbit.  These directions can be
nonlinearly extended into the unstable and stable manifolds of the periodic
orbit.  A sufficient condition for the periodic orbit to be hyperbolic is that
its Floquet matrix have nondegenerate eigenvalues.  (The Floquet matrix is
obtained by linearising the system about the periodic orbit and integrating
over a full period of the orbit, as we will do in Section~\ref{sec:sineflow}
for a specific flow.)  Even though they are topological obstacles to material
lines in the flow, such orbits can hardly be called ghost rods.  This is
because material lines must align with the unstable manifold of the periodic
orbit, a phenomenon sometimes referred to as asymptotic directionality
\citep{Giona1998b,Thiffeault2004c}.  A material line cannot possibly fold
around such a periodic orbit, since the unstable manifold goes straight
through the orbit and appears linear in its neighbourhood.  Hence, the
periodic orbit does not ``look'' like a tiny little rod to the naked eye: it
looks like any other point on the material line, and only a detailed knowledge
of the velocity field allows its detection, usually by numerical means.  We
conclude that hyperbolic periodic orbits do not form ``proper'' ghost rods,
since they cannot be detected visually.

That leaves the second type of unstable periodic orbit: parabolic orbits.  In
that case the Floquet matrix has degenerate eigenvalue that must both be equal
to unity, by incompressibility of the fluid.  For parabolic points, we cannot
deduce the behaviour of points near the periodic orbit by examining only the
linearised system---nonlinear terms must be considered.  As we shall see in
the following section, a particular type of nonlinear structure gives rise to
parabolic points that exhibit the appropriate behaviour for a ghost rod.

\CCLsection{Case Study: The Sine Flow}
\label{sec:sineflow}

\begin{figure}[htbp]
\centering
\subfigure[]{
\hspace{-4em}\includegraphics[width=.65\textwidth]{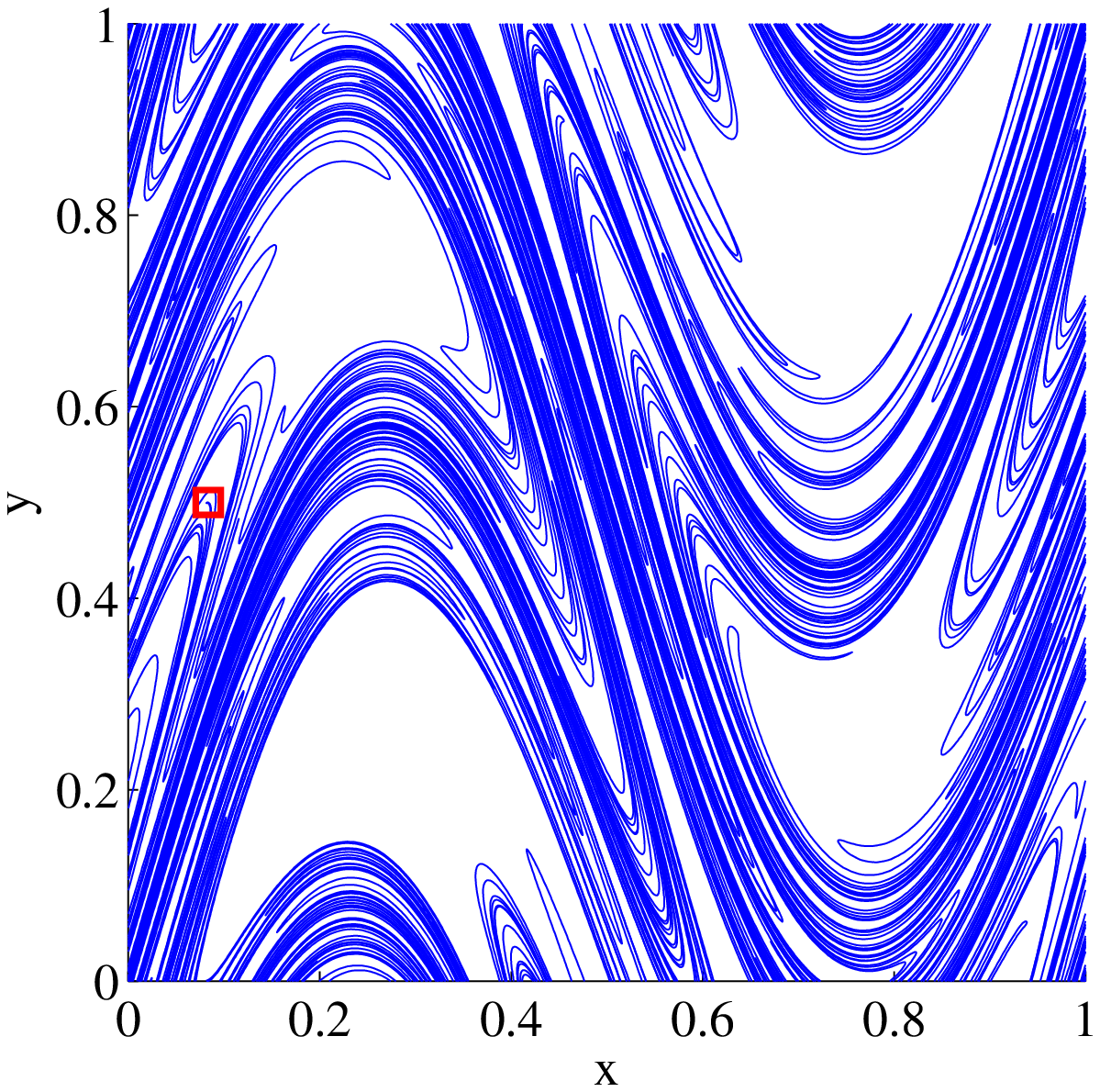}
\label{fig:sf_line}
}
\subfigure[]{
\hspace{-3em}\includegraphics[width=.5\textwidth]{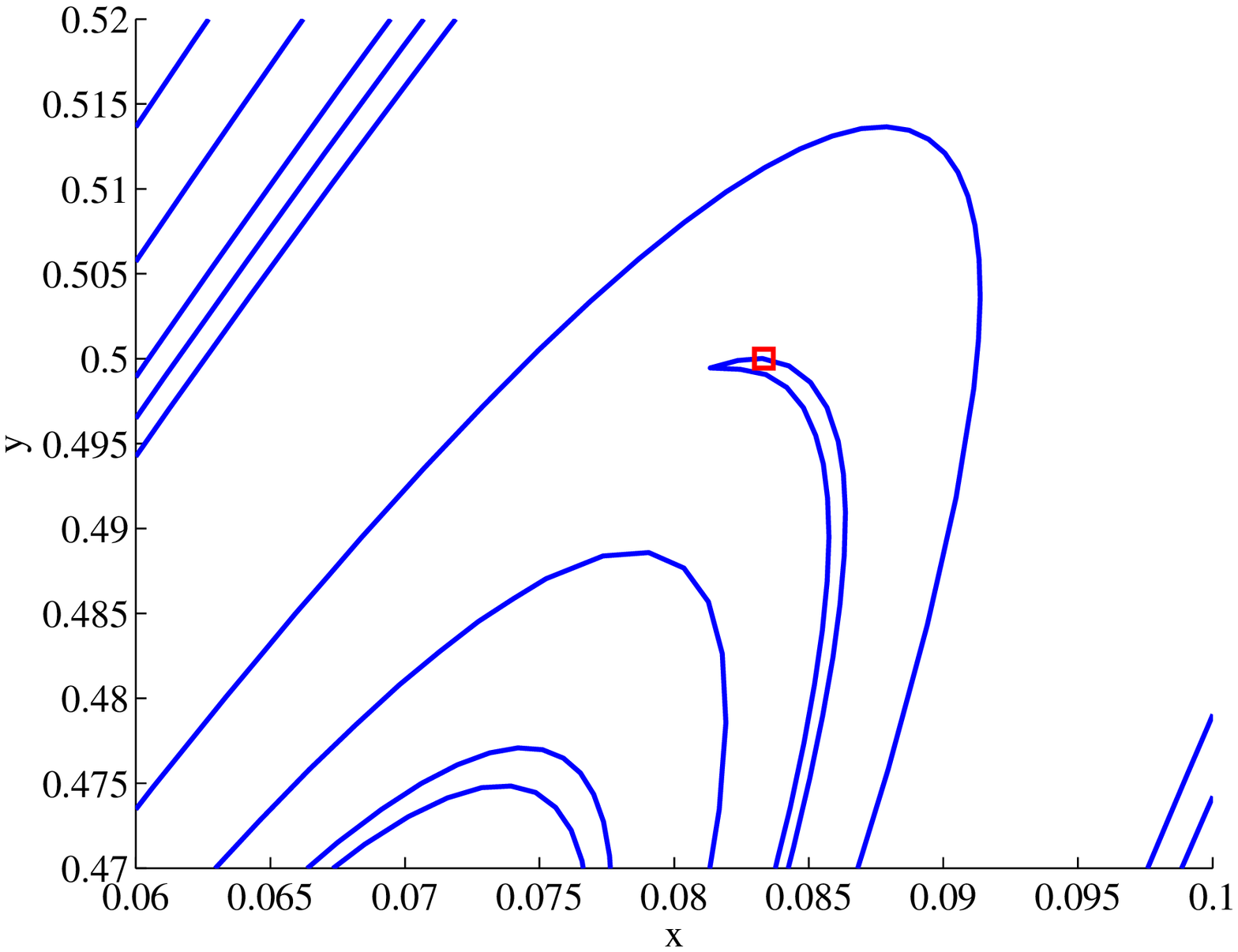}
\label{fig:sf_tip}
}
\caption{(a) A material line stretched and folded by the sine flow.  The
  parabolic periodic point at~$(x,y)=(1/12,1/2)$ is shown boxed.  (b) Blow-up
  of the periodic point.  Note that the material line is folded near, but not
  quite tightly around, the parabolic fixed point.}
\label{fig:sf_matline}
\end{figure}

We shall now illustrate the type of unstable periodic orbit that gives rise to
ghost rods by examining a specific system, namely the Zeldovich sine flow
\citep{Pierrehumbert1994}.  This is a nice system to work with because its
Poincar\'e map can be obtained analytically, and for special parameter values
we can also determine some periodic orbits analytically.  These orbits were
exploited by \cite{Finn2005} to show the presence of chaos.  The sine flow is
given by the velocity field
\begin{equation}
  \velv(\xv,\time) =
  \begin{cases}
    (0\,,\,\sin 2\pi\xc), \quad & \iter\tau \le \time <
    (\iter+\tfrac{1}{2})\tau;\\
    (\sin 2\pi\yc\,,\,0), \quad & (\iter+\tfrac{1}{2})\tau \le \time <
    (\iter+1)\tau,
  \end{cases}
\end{equation}
where~$\iter$ is an integer.  The equation~$\dot\xv=\velv(\xv,\time)$ can then
be integrated over one period to give the sine map
\begin{equation}
\begin{split}
\xcii &= \xci + \tfrac{1}{2}\,\tau\,\sin 2\pi\ycii\,;\\
\ycii &= \yci + \tfrac{1}{2}\,\tau\,\sin 2\pi\xci\,,
\end{split}
\label{eq:sinemap}
\end{equation}
with~$\xv_\iter \ldef \xv(\iter\tau)$.  As an example, we will take~$\tau=1$,
because then we can determine many periodic orbits analytically.  For
instance, there is a period-4 orbit starting at~$\xv_0=(1/12,1/2)$, with
iterates
\begin{equation}
  (1/12,1/2) \rightarrow (7/12,3/4) \rightarrow
  (7/12,1/2) \rightarrow (1/12,1/4) \rightarrow (1/12,1/2).
  \label{eq:po}
\end{equation}
The initial location of this orbit is inside the small square in
Fig.~\ref{fig:sf_line}, which also shows a material line advected for a few
periods of the sine flow.
Figure~\ref{fig:sf_tip} is a blow-up of the material line near this periodic
orbit.  Notice how the material line is sharply folded around the periodic
orbit.  In fact, Fig.~\ref{fig:sf_line} contains several such sharp folds.
They are quite generic in chaotic flows, and are associated with regions of
anomalously low stretching \citep{Liu1996,Thiffeault2004c}.

We are interested in the behaviour of the map~\eqref{eq:sinemap} in the
neighbourhood of this periodic orbit, so we define a
variable~$\tXv\ldef\xv-\xv_0$.  Then, we can expand the map to second order
in~$\tXv$,
\begin{equation}
\begin{split}
  \tXc' &= \tXc - 2\pi\tYc + \qfpar_{\tXc}\,\tXc^2
  + \beta_{\tXc}\,\tYc^2 + \gamma_{\tXc}\,\tXc\tYc,\\
  \tYc' &= \tYc \phantom{- 2\pi\tYc} \ \, + \qfpar_{\tYc}\,\tXc^2
  + \beta_{\tYc}\,\tYc^2 + \gamma_{\tYc}\,\tXc\tYc,
\end{split}
\label{eq:quadmapt}%
\end{equation}%
where the primes denote four iterations of the sine map, so that the periodic
orbit has become a fixed point of the map~\eqref{eq:quadmapt} at~$\tXv=(0,0)$.
The periodic orbit is parabolic, as can easily be seen from the fact that the
linear part of~\eqref{eq:quadmapt} (the Floquet matrix) has matrix
representation
\begin{equation}
  \Jac = \begin{pmatrix}1 & -2\pi \\ 0 & 1\end{pmatrix},
    \label{eq:Jordan}
\end{equation}
which implies unit eigenvalues for the map.  However, this matrix is not
diagonalisable: it only has one eigenvector,~$(0\ \ 1)^T$ (this can only occur
for a parabolic point).  We will see that it is this nondiagonalisable nature
that allows the ``folding'' of material lines around the periodic point.  In
general, a matrix~$\Jac$ has this property if~$(\Jac - \mathbb{I})^2=0$,
for~$\Jac\ne\mathbb{I}$, which given that~$\det\Jac=1$ is equivalent
to~$\trace\Jac=2$, with~$\Jac\ne\mathbb{I}$.

After a linear transformation and a near-identity area-preserving quadratic
transformation, Eq.~\ref{eq:quadmapt} can be brought into the form
\begin{equation}
\begin{split}
  \Xc' &= \Xc + \Yc + \qfpar\,\Xc\Yc\,,\\
  \Yc' &= \Yc + \qfpar\l(\tfrac{1}{2}\Xc^2 + \Xc\Yc\r),
\end{split}
\label{eq:quadmap}%
\end{equation}%
where the coefficients are such that the map is area-preserving to linear
order.  (The transformation used to get to~\eqref{eq:quadmap} is not generally
orientation-preserving.)  As long as the linear part of the system is a Jordan
block of the form~\eqref{eq:Jordan}, we can transform the system to
Eq.~\eqref{eq:quadmap}.  We have thus reduced the dynamics near the parabolic
point to a one-parameter map (basically a H\'enon map), which we proceed to
analyse.

\CCLsubsection{Invariant Manifolds and Dynamics Near the Origin}

We now want to find the shape of the unstable and stable manifolds of the
fixed point of~\eqref{eq:quadmap} at the origin.  Unlike hyperbolic fixed
points, for a parabolic point the invariant manifolds cannot be determined
solely from the linear part.  Rather, we use the invariance property of the
manifold: we parametrise the invariant manifold by~$(\Xc,\Yci(\Xc))$ and
iterate the map,
\begin{subequations}
\begin{align}
  \Xc' &= \Xc + \Yci(\Xc) + \qfpar\,\Xc\Yci(\Xc)\,,
  \label{eq:quadmapinva}\\
  \Yci(\Xc') &= \Yci(\Xc) + \qfpar\l(\tfrac{1}{2}\Xc^2 + \Xc\Yci(\Xc)\r),
  \label{eq:quadmapinvb}
\end{align}
\label{eq:quadmapinv}%
\end{subequations}%
where we wrote~$\Yc'=\Yci(\Xc')$ since, by the invariance property, the
iterated point must still belong to the invariant manifold.  We can then
substitute~\eqref{eq:quadmapinva} into~\eqref{eq:quadmapinvb},
\begin{equation}
  \Yci(\Xc + \Yci(\Xc) + \qfpar\,\Xc\Yci(\Xc))
  = \Yci(\Xc) + \qfpar\l(\tfrac{1}{2}\Xc^2 + \Xc\Yci(\Xc)\r),
\end{equation}
which is an equation that must be solved for~$\Yci(\Xc)$.  We are interested
in the small~$\Xc$ form of the manifold, so we assume~$\Yci(\Xc) =
\sigma\Xc^\delta$ and try to balance the leading order terms:
\begin{equation}
  \sigma\,(\Xc + \sigma\Xc^\delta + \qfpar\,\sigma\Xc^{1+\delta})^\delta
  = \sigma\Xc^\delta + \qfpar\l(\tfrac{1}{2}\Xc^2 + \sigma\Xc^{1+\delta}\r).
\end{equation}
Where we go next depends on the magnitude of~$\delta$.  If~$\delta=1$, we get
the equation~$\sigma(1+\sigma)=\sigma$ for the coefficients of the linear
terms, which implies~$\sigma=0$, an unacceptable state of affairs since then
the quadratic term is unbalanced.  If~$\delta<1$, we get the leading balance
\begin{equation}
  \sigma\,(\sigma\Xc^\delta)^\delta
  = \sigma\Xc^\delta.
\end{equation}
This can only be satisfied for~$\delta=1$, a contradiction, or~$\delta=0$,
which again leads to an unbalanced quadratic term.  Hence, our only choice is
to take~$\delta>1$, which gives the leading-order balance
\begin{equation}
  \sigma\,(\Xc + \sigma\Xc^\delta)^\delta
  = \sigma\Xc^\delta + \tfrac{1}{2}\qfpar\Xc^2,
\end{equation}
where we have kept an extra order, because the leading terms cancel after
expanding the exponent,
\begin{equation}
  \sigma\Xc^\delta\,(1 + \sigma\delta\Xc^{\delta-1})
  = \sigma\Xc^\delta + \tfrac{1}{2}\qfpar\Xc^2,
\end{equation}
and we get finally
\begin{equation}
  \sigma^2\,\delta\Xc^{2\delta-1}
  = \tfrac{1}{2}\qfpar\Xc^2,
\end{equation}
yielding~$\delta=3/2$, $\sigma=\pm\sqrt{\qfpar/3}$.  We can thus write the
shape of the invariant manifolds as
\begin{equation}
  \Yci(\Xc) = \pm\sqrt{\tfrac{1}{3}\,\qfpar\Xc}\Xc\,,
  \label{eq:invmanif}
\end{equation}
to leading order, where we have written~$\qfpar\Xc$ under the square root to
show that~$\Xc$ and~$\qfpar$ must have the same sign.  Note an important fact:
the manifolds exist only on one side of the~$\Xc$ axis, in contrast to
Fig.~\ref{fig:hyper} where the manifolds must radiate from the fixed point in
four directions.

The two solutions for~$\sigma$ correspond to the stable and unstable
manifolds.  We can determine which sign goes with which manifold by looking at
an iterate of~$\Xc$ in Eq.~\eqref{eq:quadmapinva},
\begin{equation}
  \Xc' = \Xc \pm\sqrt{\tfrac{1}{3}\,\qfpar\Xc}\Xc\,,
\end{equation}
where we neglected higher-order terms ($\Xc^{5/2}$).  Thus, for~$\qfpar>0$,
which implies~$\Xc>0$, the~`$+$' solution takes~$\Xc$ farther from the origin
(unstable manifold), whilst the~`$-$' solutions takes the point closer to the
origin (stable manifold).  The situation is reversed for~$\qfpar<0$.

\begin{figure}[htbp]
\psfrag{it=0}{$\scriptstyle \iter=0$}
\psfrag{it=7}{$\scriptstyle \iter=7$}
\psfrag{it=14}{$\scriptstyle \iter=14$}
\psfrag{it=21}{$\scriptstyle \iter=21$}
\psfrag{it=28}{$\scriptstyle \iter=28$}
\psfrag{it=35}{$\scriptstyle \iter=35$}
\psfrag{it=42}{$\scriptstyle \iter=42$}
\psfrag{it=49}{$\scriptstyle \iter=49$}
\psfrag{it=56}{$\scriptstyle \iter=56$}
\psfrag{X}{$\Xc$}
\psfrag{Y}{$\Yc$}
\centering
\includegraphics[width=\textwidth]{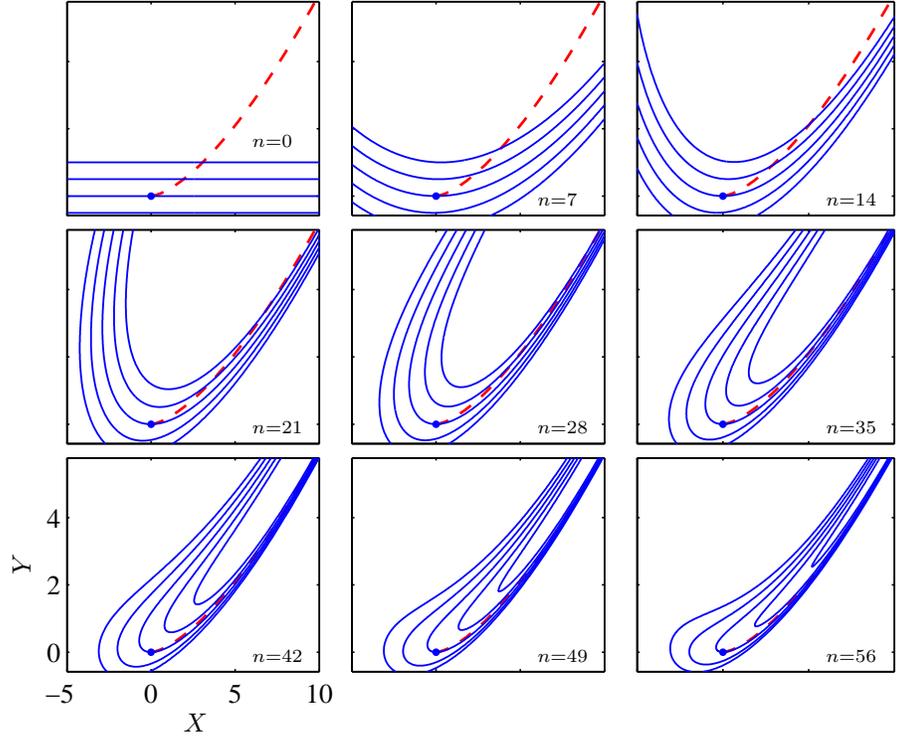}
\caption{Iteration of a few material lines by the map~\eqref{eq:quadmap}
for~$\qfpar=1$.  The lines fold around the unstable manifold (dashed
curve). The~$\Xc$ axis has been rescaled by~$10^{-3}$, the~$\Yc$ axis
by~$10^{-4}$.}
\label{fig:folditer}
\end{figure}

Figure~\ref{fig:folditer} shows a few iterates of horizontal lines
under the action of the map~\eqref{eq:quadmap}.  The linear part of the map
acts as a ``shear flow'' that sweeps the line around the origin, but the
nonlinear terms prevent the line from crossing the unstable manifold.  The net
result is a line folded around the unstable manifold.  This is why it is
appropriate to refer to~\eqref{eq:quadmap} as the `folding normal form':
inside every sharp fold of the flow lurks such a map.\footnote{At the
workshop, Stefano Cerbelli and Massimiliano Giona pointed out that their
research also seems to support this.} Since nearby material lines align with
the unstable manifold of the periodic orbit, the folding is made possible by
the one-sidedness of the unstable manifold: unlike hyperbolic points
(Fig.~\ref{fig:hyper}), the manifold does not traverse the parabolic periodic
point, but instead terminates there.  This allows material lines to wrap
around the periodic point without encountering the invariant unstable
manifold, which cannot be crossed.

\CCLsubsection{Curvature}
\label{sec:curv}

As time progresses the folds in the material lines in Fig.~\ref{fig:folditer}
come closer and closer to the periodic orbit.  There seems to be no limit to
how close they can come, which is consistent with the ghost rod having zero
effective size.  The best we can do is to characterise the effective size of
the ghost rod by how quickly the curvature of the folds evolve.  We shall now
examine how the curvature of a material line evolves near the parabolic orbit.

\begin{figure}[htbp]
\psfrag{slope = \(3/2\)sqrt\(3\)pi5}{$\text{slope} = (3/2)\sqrt{3}\pi^5$}
\psfrag{period}{iterations}
\psfrag{curvature}{curvature}
\centering
\includegraphics[width=10cm]{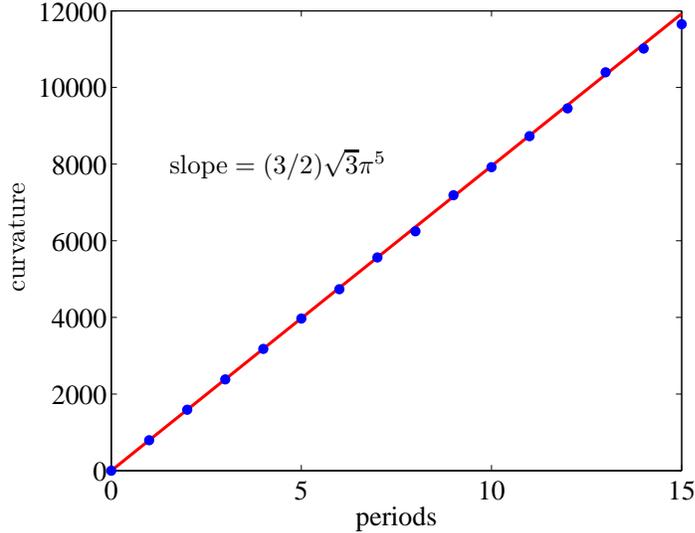}
\caption{The evolution of the curvature of a material line passing through the
  periodic point shown in Fig.~\ref{fig:sf_matline}, with orbit given by
  Eq.~\eqref{eq:po}.  The curvature at the point increases linearly with the
  number of periods, showing that the line is getting more tightly folded
  around the parabolic point.}
\label{fig:sf_tip_curv}
\end{figure}

Consider a material line going through the origin of~\eqref{eq:quadmap}, as
depicted in Fig.~\ref{fig:folditer}.  The tangent map of~\eqref{eq:quadmap}
at the origin tells us how the tangent to the curve evolves,
\begin{equation}
  \begin{pmatrix}\delta\Xc' \\ \delta\Yc'\end{pmatrix}
    = \begin{pmatrix}1 & 1 \\ 0 & 1\end{pmatrix}
  \begin{pmatrix}\delta\Xc \\ \delta\Yc\end{pmatrix},
    \label{eq:linsys}
\end{equation}
where~$(\delta\Xc\ \ \delta\Yc)^T$ is the tangent.  The second variation
of~\eqref{eq:quadmap} is
\begin{equation}
  \begin{pmatrix}\delta^2\Xc' \\ \delta^2\Yc'\end{pmatrix}
    = \begin{pmatrix}1 & 1 \\ 0 & 1\end{pmatrix}
  \begin{pmatrix}\delta^2\Xc \\ \delta^2\Yc\end{pmatrix}
    + \qfpar\begin{pmatrix}\delta\Xc\delta\Yc \\
    (\delta\Xc)^2 + \delta\Xc\delta\Yc\end{pmatrix}.
    \label{eq:quadsys}
\end{equation}
For the case shown in Fig.~\ref{fig:folditer}, the initial tangent is
parallel to~$(1\ \ 0)^T$, which is an eigenvector of the matrix
in~\eqref{eq:linsys}: the tangent doesn't change.  Given that~$\delta^2\Xc$
and~$\delta^2\Yc$ are zero initially (the line is straight), we can
solve~\eqref{eq:quadsys} for~$\delta^2\Yc$,
\begin{equation}
  \delta^2\Yc = \iter\,\qfpar(\delta\Xc)^2\,,
  \label{eq:d2Ysol}
\end{equation}
where~$\iter$ is the number of iterations.  The curvature of the line is given
by \citep{Liu1996}
\begin{equation}
  \kappa 
  = \frac{\delta\Xc\delta^2\Yc - \delta\Yc\delta^2\Xc}
  {\l\lVert\delta\Xv\r\rVert^3}\,.
\end{equation}
Now given the solution~\eqref{eq:d2Ysol} and the fact that~$\delta\Yc=0$ for
all time, the curvature evolves as
\begin{equation}
  \kappa = \iter\,\qfpar\,,
  \label{eq:curvevol}
\end{equation}
so the curvature of the material line grows linearly with time.  This is
verified by a calculation with the sine flow, for the periodic
orbit~\eqref{eq:po}: Figure~\ref{fig:sf_tip_curv} shows that the curvature of
a material line anchored at the periodic orbit does indeed grow linearly with
the number of iterations, and the slope is in perfect agreement with the
results from the folding normal form.

Note that this linear evolution of the curvature is not an artefact of our
choice of orientation of the material line.  If we choose the line to be
orthogonal to the one in Fig.~\ref{fig:folditer}, we find that the tangent
evolves as~$(\delta\Xc\ \ \delta\Yc)^T = (\iter\ \ 1)^T\delta\Yc_0$, which
means that the tangent aligns with the direction of the unstable manifold for
large~$\iter$.  The curvature will then grow asymptotically at the rate given
by~\eqref{eq:curvevol}.

It is clear from Fig.~\ref{fig:folditer} that the point of highest curvature
is not at origin.  Nevertheless, the increase in curvature is linear in the
neighbourhood of the periodic orbit, and all material lines near the orbit
eventually wrap around it, so the orbit can still be said to be acting as a
ghost rod.

We conclude that $\qfpar$ measures the ``size'' of the rod: a higher~$\qfpar$
means that material lines converge towards the periodic orbit more rapidly, so
that the ghost rod has a smaller apparent impact on the flow.
Visually,~$\qfpar$ could be estimated by looking at the rate at which material
lines ``bunch up'' near a periodic orbit, as in Fig.~\ref{fig:folditer}, but
in practice this is quite difficult.

\CCLsection{Discussion}
\label{sec:discussion}

Of course, this paper is just a sketch of a theory for the size of ghost rods:
a comprehensive theory remains to be developed.  Rather, we tried to give an
indication of the factors that influence a ghost rod's apparent size.

The motivation behind this study, and the ghost rod framework in general, is
to try to determine some stirring properties of a flow from visual cues.  It
is obvious that we can determine the size of physical rods by just looking at
them.  The effective size of ghost rods that are elliptic islands can also be
determined visually, as is apparent in Fig.~\ref{fig:ghostrod}.  Ghost rods
associated with parabolic points are harder to identify: they typically occur
inside sharp folds in material lines, as in Fig.~\ref{fig:sf_matline}.  Even
if they are identified, measuring their effective impact on the flow is far
from trivial: one can attempt to measure the evolution of curvature near the
point, in the same spirit as in Section~\ref{sec:curv}, or see how rapidly
material lines bunch-up near the periodic orbit, effectively a measure of the
coefficient~$\qfpar$.

The sharp folds observed in material lines are the spots where the stretching
is weakest, because there is usually a competition between stretching and
curvature \citep{Liu1996,Thiffeault2004c}.  Hence, folds are associated with
inhomogeneities in the stretching field, and thus typically decrease the
efficiency of stirring since uniformity is desirable.  Knowing how fast the
curvature grows (as measured by~$\qfpar$) gives a hint of how much
inhomogeneity a fold introduces, and thus quantifies its impact on the quality
of mixing.

The parabolic points may be the ``relevant'' ghost rods, \ie\ the ones on
which one can construct a braid that captures exactly the topological entropy
of the flow \citep{Gouillart2005preprint}.  We have no proof of this
assertion yet; however, since the folds determine the skeleton around which a
material line will wrap, these points certainly play a distinguished role.

\CCLsection*{Acknowledgments}

We thank the organisers, Luca Cortelezzi and Igor Mezi\'{c}, as well as
Stefano Cerbelli and Massimiliano Giona for helpful discussions during the
workshop.  This work was funded by the UK Engineering and Physical Sciences
Research Council grant GR/S72931/01.


\end{document}